\documentclass{emulateapj}
\usepackage{apjfonts}
\usepackage{graphicx}

\def\etal  {{\it et al.}\ }
\def\kms{${\rm km~s^{-1}}$}

\def\spose#1{\hbox to 0pt{#1\hss}}
\def\lta{\mathrel{\spose{\lower 3pt\hbox{$\mathchar"218$}}
     \raise 2.0pt\hbox{$\mathchar"13C$}}}
\def\gta{\mathrel{\spose{\lower 3pt\hbox{$\mathchar"218$}}
     \raise 2.0pt\hbox{$\mathchar"13E$}}}

\slugcomment{11 Oct  2005 Accepted for ApJ Letters}

\shorttitle{Colors of Field \& Cluster Bulges}
\shortauthors{Koo et al.}

\begin{document}

\title{Colors of Luminous Bulges in Cluster MS1054-03 and Field Galaxies at
Redshifts $z \sim 0.83$\altaffilmark{1}}


\author{David C. Koo\altaffilmark{2},
Susmita Datta\altaffilmark{2},
Christopher N. A. Willmer\altaffilmark{2,3},
Luc Simard\altaffilmark{2,4}, 
Kim-Vy Tran\altaffilmark{5}, 
and Myungshin Im\altaffilmark{6},
}

\altaffiltext{1} {Based on observations obtained at the W. M. Keck
Observatory, which is operated jointly by the University of California
and the California Institute of Technology; obtained 
with the NASA/ESA {\it Hubble Space Telescope (HST)} from the
data Archive at the Space Telescope Science Institute (STScI), which
is operated by the Association of Universities for Research in
Astronomy, Inc., under NASA contract NAS5-26555; 
and associated with proposals GO 7372, GTO 5090, and GTO 5109.}

\altaffiltext{2} {UCO/Lick Observatory, Department of Astronomy and 
Astrophysics, University of California, Santa Cruz, CA 95064; 
Electronic Mail: koo@ucolick.org, cnaw@ucolick.org, susmita@ucolick.org}

\altaffiltext{3} {On leave from Observatorio Nacional,MCT, CNPq, Rio de Janeiro, Brazil}

\altaffiltext{4} {Herzberg Institute of Astrophysics, National Research Council of Canada, 5071 West Saanich Road, Victoria, BC, V9E 2E7 
Canada; Electronic Mail: Luc.Simard@hrc.ca}

\altaffiltext{5} {Institute of Astronomy, ETH Z\"urich, CH-8093 Z\"urich, 
Switzerland; Electronic Mail: vy@phys.ethz.ch}

\altaffiltext{6} {Astronomy Program, SEES, Seoul National University, Seoul, Korea}


\newpage

\begin{abstract}

Using $HST$ images, we separate the bulge-like (dubbed pbulge) and
disk-like (pdisk) components of 71 galaxies in the rich cluster
MS1054-03 and of 21 in the field.  Our key finding is that luminous
pbulges are very red with restframe $U-B \sim 0.45$, while predicted
colors are bluer by 0.20 mag.  Moreover, these very red colors appear
to be independent of environment, pbulge luminosity, pdisk color, and
pbulge fraction.  These results challenge any models of hierarchical
galaxy formation that predict the colors of distant ($z \sim 0.8$)
luminous field and cluster bulges would differ.  Our findings also
disagree with other claims that 30\% to 50\% of bright bulges and
ellipticals at $z \sim 1$ are very blue ($U-B \leq 0$).
\end{abstract}



\newpage


\section{Introduction}
The ages of bulges (also commonly referred to as spheroids)\footnote
{defined here to include {\it ellipticals} and the bulges of S0's and
spirals.  E, E/SO, and S0 morphological types together as a class are
often called spheroidal galaxies to be designated in this paper as
E-S0, to avoid the ambiguity of the transition-class E/S0 designation
and confusion with bulges or spheroids.} remain an important unsolved
problem in stellar populations and galaxy formation (Wyse, Gilmore, \&
Franx 1997). The three current and contending paradigms are early,
rapid single-collapse formation of bulges and ellipticals at redshifts
$z > 2$; formation at later epochs via merging of disk galaxies; and
the formation of bulges through secular evolution processes in disks
(see review by Kormendy \& Kennicutt 2004).  In general, distant
cluster galaxies show consistent results from different studies,
namely a tight color-magnitude relation for E-S0 and very red colors
that persist to redshifts $z \sim 1$ (see, e.g., van Dokkum \etal
2000, Blakeslee \etal 2003, Menanteau \etal 2004).  Both support the
early bulge formation scenario.  In contrast, field\footnote {field
refers to regions outside of rich galaxy clusters} E-S0 and bulges
show inconsistent and inconclusive results.  For example, using the
luminosity function as a test between the two scenarios, Lilly \etal
(1996) find little luminosity or density evolution for red galaxies;
Im \etal (2002 [GSS10]) see significant luminosity evolution and
little volume density evolution of morphologically selected E-S0 to $z
\sim 1$; and Bell \etal (2004) find luminosity and density evolution
of red galaxies (presumed to be dominated by E-S0 galaxies) that
maintain a constant luminosity density with lookback time to $z \sim
1$ (see recent review by McCarthy 2004).  
Perhaps the picture is mixed for galaxies with different masses, as
suggested by fundamental plane studies at these redshifts 
showing less massive galaxies
appearing to be younger (e.g., Treu \etal 2005).
Using colors as a surrogate
for estimated ages, some studies support the merger picture by finding
or claiming a large fraction (30\% to 50\%) of field bulges and/or 
early-type galaxies at
intermediate redshifts to be blue or to show evidence of recent star
formation (e.g., Kodama, Bower, \& Bell 1999;
Abraham \etal 1999; Schade \etal 1999; Tamura \etal 2000; Menanteau
\etal 2001 \& 2004; Ellis, Abraham, \& Dickinson 2001), while a few other
surveys find uniformly red E-S0 (GSS10) or bulges (Koo \etal 2005
[GSS8]) at the same epochs.

As shown by Menanteau \etal (2001, 2004), the properties of field
bulges versus cluster bulges provide powerful diagnostics of formation
mechanisms.  They found, contrary to expectations of hierarchical
formation models, that the bulges of field spirals appeared bluer on
average than field ellipticals (c.f., GSS8 where both are found to be
the same). Using cluster ellipticals as a benchmark for old stellar
populations, Menanteau \etal (2001) estimated that ``at $z \sim 1$
about half the field spheroidals must be undergoing recent episodes of
star formation.''  Moreover, as emphasized by the work of Benson \etal
(2002), the resolved colors of spheroids are particularly effective
probes of their formation history.

Based on decomposing galaxies into bulge-like and disk-like
components, Figure 5 in GSS8 shows that the luminous field pbulges are
{\it redder} and have a {\it shallower} color-magnitude relation (CMR)
slope than that derived from {\it integrated colors} of early-type
{\it galaxies} in the MS1054 cluster. Without removing the
contamination by potentially bluer disks, we do not yet know whether
cluster {\it bulges} are indeed bluer than field or have a steeper CMR
slope.  Moreover, the bulge colors of the cluster spirals, i.e.,
non-early-type, have yet to be measured. Photometry that separates
bulges from disks of cluster galaxies is clearly needed.  This letter
presents the results of such a study.

We adopt a Hubble constant H$_o$ = 70~\kms Mpc$^{-1}$ and a flat
cosmology with $\Omega_m = 0.3$ and $\Omega_{\Lambda} = 0.7$.  Our
photometry is in the Vega system with $V$ referring to $F_{606}$ and
$I$ to $F_{814}$ for the WFPC2 data from $HST$.  At redshift $z \sim
0.83$, $V-I$ is very close to restframe $U-B$ and $I$ is close to
restframe $B$, thus making any K-corrections to these measures very
small. Observed colors of $V-I \sim 2.2$ are close to restframe $U-B
\sim 0.45$ and,  for such very red stellar populations, $I \sim 22$ is
equivalent to blue absolute magnitudes of $M_B \sim -20.56$.

\section{Observations} \label{obs}

Cluster MS1054-03 at redshift $z \sim 0.83$ was chosen for analysis
since it is among the best observed distant clusters. The cluster has
relatively deep (6600s in each of $V$ and $I$) and extensive (6
pointings of WFPC2 covering about 30 $\sq'$ of the region) $HST$
imaging from HST program GO 7372 (see Hoekstra et al. 2000, van Dokkum
et al. 2000). It also possesses over 100 spectroscopic redshifts (Tran
\etal 2005, in preparation).  Flat fielding and cosmic-ray removal of
the $HST$ images are described in detail by van Dokkum \etal (2001)
who also measured colors and magnitudes via aperture photometry for 81
galaxies with spectroscopic redshifts (van Dokkum \etal 2000).

Since our goal is to isolate the bulge component, the reduced $HST \
V$ and $I$ images were analyzed with our own photometry package
(GIM2D: see Simard \etal 2002[GSS2] and GSS8 for details) to separate
the luminosities and colors of the bulges and disks. GIM2D fits each
object in an image with a PSF-convolved 2D two-component (bulge+disk)
model. The two components are 1) the bulge, which is assumed to follow
an $r^{1/4}$ profile (de Vaucouleurs profile) that characterizes many
luminous ellipticals and bulges (Andredakis, Peletier, \& Balcells
1995), but this is a simplification (see de Jong \etal 2004 for a more
in-depth study of local cluster bulge profile shapes using GIM2D), and
2) the disk, which is fit with an exponential profile. Since we cannot
be sure that the components extracted from the decomposition are
genuine bulges and disks, we referred to these measured components as
photometrically-derived bulges and disks or photobulge and photodisk
in GSS8.  In this Letter, we dub these as pbulge and pdisk.

Previous work with GIM2D (GSS2, GSS8) have shown that reliable
two-band colors for compact bulges are derived by simultaneous fits of
the profile parameters and the adoption of the simplifying assumption
that color gradients for either of the subcomponents are small. One
image in each filter is used in GIM2D. The pbulge and pdisk centers,
sizes, orientation, and shapes are constrained to be the same for the
two filters, while the pbulge-to-total ratio and total luminosity are
left to float. Undersampled WFPC2 images with sub-pixel shifts cannot
be re-sampled and stacked without loss of information, due to
aliasing, unless a large number of images are available for stacking.
Given that this is not case for our data, we make use, instead, of
GIM2D's option of simultaneously fitting a series of non-rotated
images in a single filter, each with its own set of PSF's. However, we
cannot yet undertake simultaneous fits with multiple images in
multiple filters.

The cluster has 3 pairs of offset images in each of $V$ and $I$.  As a
workaround to the limitations of GIM2D, we first fit each object with
the 3 pairs of $I$ images.  The $I$ was chosen since the bulges were
generally very red and thus had higher S/N in $I$ than in $V$.  Except
for the luminosity and pbulge-to-total ratio, all the other measured
parameters from $I$ were locked when the $V$ image was fit.  The
conversion of the measured $I$ and $V-I$ to restframe $M_B$ and $U-B$
were made using the method described by Weiner \etal (2005: GSS3) and
Willmer \etal (2005), but yield values very close to that adopted by
van Dokkum \etal (2000). Color conversion errors can be seen to range
from 0.05 to 0.08 mag in Fig. 12 of Gebhardt \etal (2003: GSS9).

The field bulges were chosen from a sample of galaxies within the $V$
and $I$ $HST$ images of the ``Groth Strip Survey'' (GSS), a ``chevron
strip'' of 28 WFPC2 pointings oriented NE to SW at roughly 14:17+52 at
Galactic latitude $b \sim 60\deg$ (programs GTO 5090, 5109) covering
an area of 127 arcmin$^{2}$ (Vogt et al. 2005).  The galaxies were
constrained to have about the same redshifts as MS1054-03, namely $z$
between 0.79 and 0.87, where the spectroscopic redshifts were measured
by {\it DEEP}{\footnote {Deep Extragalactic Evolutionary Probe: see
URL http://deep.ucolick.org/}}. From a total sample of over 600
redshifts (Weiner \etal 2005), 76 fell within our redshift range {\it
and} had HST imaging. Except for one ultra-deep pointing
(14:17.5+52.5, Westphal Deep Survey Field 2) with 24,400s in $V$ and
25,200s in $I$, the other HST images in GSS were comprised of a
combined image in $I$ of 4400s and in $V$ of 2800s.  As detailed in
GSS2, GIM2D processing of the 2800s $V$ and 4400s $I$ were made
separately and simultaneously.  To match the procedure for the cluster
photometry using GIM2D, we adopted and locked the parameters from the
4400s $I$ (except for luminosity and pB/T ratio) and used these to
process the multiple images in $V$.  The conversion of the input $I$
and resultant $V-I$ to restframe $M_B$ and $U-B$ were as mentioned
above.

Our prior experience with GIM2D (GSS8) suggests four additional
criteria to define the final sample, with brackets enclosing the
number of galaxies excluded by each criterion for the field and
cluster, respectively: 1) [45,33] the pbulge had to be brighter than
$I = 23$ (M$_B \sim$ -19.5) to yield reliable photometry; 2) [1,2] the
pbulge half-light sizes were larger than one third of a pixel,
i.e. 0.03 arcsec, again to assure reliable photometry; 3) [5,4] the
pbulges had to have half-light sizes smaller than the half-light sizes
of the pdisk subcomponent {\it or} the pbulges were dominant with more
than 2/3 of the total luminosity, i.e., pB/T $>$ 0.67; and 4) [1,4]
the morphology was not so distorted as to be poorly represented by our
assumed two components. The third criterion was needed to reject cases
where GIM2D fits a central probable bulge with a small exponential
component and the outer probable disk with an $r^{1/4}$
component. Such structures appear to exist in nature but we chose to
reject such anomalous bulges from this survey. Examples of rejected
galaxies based on morphology include, e.g., a cluster galaxy that
appears to be a spiral with two holes (ID 1403: see image in van
Dokkum \etal 2000) and a field galaxy that appears to be a highly
distorted close merger (GSS ID 073\_1809: see image in GSS8). 
Another that was retained looks like  a tulip with a
stem (GSS ID 174\_4356: see GSS8) and should perhaps be rejected;
it is the lone very blue outlier in Fig. 1.
Seven other blue pbulges seen in Fig. 5 of GSS8 (most are not genuine
bulges or have poor fits) lie outside the redshift range adopted here.

Moreover, three of the field pbulges from GSS8 did not converge with
our revised GIM2D procedures (GSS IDs: 143\_2770, 152\_5050, and
292\_6262).  After the restrictions, the full sample of 114 MS1054
cluster members was reduced to a final sample of 71 and the full GSS
field sample of 76 was reduced to 21 pbulges.
 
\section{Results} \label{results}
The color magnitude diagrams (CMDs) of the pbulges are displayed in
Fig. 1. To equalize the  samples, the cluster pbulges have been
segregated  by their distance from the
cluster center at J2000 10:56:59.93 -03:37:36.6 of galaxy 1484 in van Dokkum
\etal (2000). The virial radius or
$R_{200}$ measured as 241$''$ by Tran \etal (1999) or 1.83 Mpc has
been revised using a larger sample of 130 galaxies and is 216$''$ or
1.64 Mpc.  All the cluster galaxies in our sample lie well within
R$_{200}$.

The main result is clear from Fig.1: the slope, color, and dispersion
of the CMDs are similar among the four samples.  Relative to the solid
line showing the median color of $U-B = 0.438$ from the 71 cluster
pbulges, the field pbulges appear slightly redder.  We also see from
Fig. 1 that the pbulge colors are red, independent of the distance
from the cluster center (i.e., average density), pbulge to total
(pB/T) ratio, and the color of the pdisk.  We also find no support for
claims or findings (see references in introduction) of 30\% to 50\%
fractions of bulges or spheroidals being very blue ($U-B \leq 0$) in
the field at this redshift or within a broader range of redshifts of
0.7 to 1.0 (GSS8). Finally, we find that almost all pbulges are as red
or redder than pdisks in both field and cluster.

Before continuing, a few caveats and possible weaknesses of the survey
are noted. The results here apply only to 1) bulges that are luminous
($M_{B} \lta -19.5$) and thus opens the possibility that lower
luminosity bulges may have experienced a different formation history;
2) galaxies that are dominatd by one or two components, with the bulge 
having an $r^{1/4}$ profile while the disk has an
exponential; 3) galaxies that do not have large enough color gradients
in either subcomponent to change the results; and 4) bulges that do
not belong to obviously merging systems.  Our field sample with only
21 galaxies is small, so that different properties might apply to 20\%
minorities.

To quantify the CMD, Table 1 gives the results of fits to the data
shown in Fig. 1, using the biweight method of \cite{beers90}.  Since
we found no luminosity dependence of the colors, we locked the slope
to 0 and measured only the colors and dispersions.  We confirm what is
seen by eye, namely, that the cluster and field pbulges show very
similar colors, with a hint that the inner cluster pbulges may be
redder than the middle or outer cluster pbulges and that field pbulges
may be slightly redder than cluster pbulges on average, but none of
these differences are statistically significant (95\% confidence limit
or better, i.e., twice the RMS from Table 1). The color dispersions
are also similar.  Since the measured dispersions from the fits are
smaller than that expected from the GIM2D color measurement errors, no
reliable intrinsic dispersions can be estimated, though the
implication is that they must be small. We estimate intrinsic color
dispersions of less than 0.05 mag based on the variation of the
measured values and GSS8 results.

\section{Discussion \& Summary}

The main result is that the restframe colors, slope, and color
dispersion of {\it luminous} ($M_B < -19.5$) cluster and field
pbulges at redshift $z \sim 0.8$ are nearly the same. Making the
simplifying assumption that colors are a surrogate for ages, we thus
find that field and cluster pbulges are universally old ($ > 1.5$
Gyr) with no evidence that the ages of luminous pbulges are
dependent on radial distance from the cluster center or whether they
reside in the cluster or in the field, i.e.  environment. Moreover,
while we do find slightly higher proportions of galaxies with large
pbulge fractions ($pB/T > 0.4$) and with red pdisks ($U-B >
0.25$) in cluster MS1054 than in the field, we find that the very red
pbulge colors in both the field and cluster are independent of
pbulge luminosity, pB/T, as well as pdisk colors.

The universally red pbulge colors with small dispersion found here can
be contrasted with prior studies (see references in introduction)
claiming that large fractions (30\%-50\%) of distant early-type
galaxies have blue colors or large internal color dispersions that
suggest active or recent star formation.  These other studies differ
from that undertaken here, either by including the underlying disk
light in the photometry of the bulge or by including lower luminosity
bulges. Our results also disagree with models that predict cluster
bulges are older than field bulges (e.g., Kauffmann 1996)
or that bulge ages depend on bulge fraction (e.g., Kauffmann 1996) or 
on cluster-centric distance (e.g., Diaferio \etal 2001).

Our findings for distant galaxies are, however, fully consistent with
the conclusions of \citet{pel99} for a local sample of bulges in
early-type galaxies.  They derived from $HST$ WFPC2 and NICMOS images
the half-light colors of bulges in groups and the field. The colors
were similar to that of elliptical galaxies in the Coma cluster and
implied an age of 10 Gyr, corresponding to redshifts $z \sim 2$. The
color spread within an effective radius where dust extinction was
negligible was also found to be equally small ($\sim$ 0.10 mag in
$B-I$, equivalent to 0.05 mag in $U-B$), implying an age spread of at
most 2 Gyr. Finally, because the bulge ages were so similar and old,
they concluded that secular evolution of disks would have difficulty
in forming bulges in early-type galaxies. Exponential bulges of many
late-type spirals, however, appear to have colors and thus presumably
ages that are correlated with their disk colors, and lending support
to secular evolution scenarios for this class of bulges (Carollo \etal
2001; Kormendy \& Kennicutt 2004). This work places no constraint on
such bulges if they are of luminosity fainter than our limit of $M_B
\sim -19.5$.

The second key result is that the pbulge colors are very red with $U-B
\sim 0.45$. This color lies between the $U-B = 0.52$ at $M_B = -20.5$ of 
a fit to the CMD of 409 E-S0's from the RC3
(Schweizer \& Seitzer 1992) and the $U-B = 0.42$  
of 379 S0's (Table 2 of Fukugita et al 1995) or $U-B = 0.40$ of 30 
bulges in early-type spirals (Peletier \& Balcells 1996).  The rough
constancy of colors over the last 7 Gyr is a surprise for which we
have no simple, compelling explanation {\it if only single-burst
stellar populations with standard IMF's are considered}.  Since bulges
should be bluer by 0.2 mag or more in $U-B$ at lookback times
corresponding to redshifts $z \sim 0.83$, we expect to observe typical
colors nearer $U-B \sim $ 0.20 - 0.32, depending on the formation
epoch of bulges.  If, however, an old, metal-rich stellar population
were to be combined with small amounts of additional star-formation
(or equally old but bluer stars of lower metallicity), significant
changes in luminosity from passive evolution of the stellar population
can be accompanied by nearly constant restframe $U-B$ colors (GSS8,
GSS9; Schiavon \etal in preparation; Harker \etal in preparation).
Such "drizzling" or "frosting" pictures find support ranging from
detailed studies of local old stellar populations (Trager \etal 2000,
Yi \etal 2005) as well as observations that distant early-type
galaxies sometimes exhibit [OII] emission lines that may indicate
small ($\sim 1 \ M_{\odot}$yr$^{-1}$ or less), but significant amounts
of star formation (Schade \etal 1999; Willis \etal 2002; van Dokkum \&
Ellis 2003; GSS8; Treu \etal 2005). 
Since cluster early-type galaxies appear "red and
dead", with little  evidence for any residual star formation, the above
explanation would not apply. If verified, the small change in cluster
bulge colors favors, instead, stellar population solutions, such as a
truncated IMF (Broadhurst \& Bouwens 2000).

   In summary, our key result is that distant luminous pbulges at
redshift $z \sim 0.83$ have unusually red colors and that such red
colors appear to be independent of environment (cluster, field, local
density), pbulge or galaxy luminosity, pdisk color, or
pbulge fraction (pB/T). Our experiment uses only 21 field and 71
cluster galaxies at one redshift and uses one color. This initial
foray and its conclusions need to be checked by expanding the sample
sizes, broadening the range of colors, and deepening the samples. The
new surveys should aim to cover lower and higher redshifts for both
field and cluster galaxies; reach lower luminosity bulges especially
within later-type spirals; and achieve sufficient photometric
precision to positively detect the intrinsic color dispersions. These
experiments are all now possible with the new, rich, and plentiful
$HST$ ACS data in the GOODS fields (Giavalisco et al. 2004), Hubble
Ultra Deep Field (UDF: Beckwith \etal, in preparation), and many
distant rich clusters (e.g., by the ACS GTO team). With the advent of
laser guide star adaptive optics on 8-10m class telescopes, stellar
population studies of small bulges may now also be extended into the
near-infrared (thus gaining better discrimination among age,
metallicity, and extinction) with the same spatial resolution of 0.05
arcsec as with $HST$ in the optical (e.g., Melbourne \etal 2005).

\acknowledgments

We thank A. Metevier, J. Harker, G. Illingworth, P. van Dokkum, and K.
Gebhardt for their help and useful discussions and two anonymous
referees for constructive criticisms.  This work was supported in part
by NASA STScI grants HST AR-0831.01 and HST AR-09204.01-A and NSF
grant AST-0071198. M. Im was supported by grant 
No. R01-2005-000-10610-0 from the Basic Research Program of the Korea
Science \& Engineering Foundation and acknowledges the support from
the start-up fund provided by Seoul National University.


\newpage




\newpage

\begin{deluxetable}{l r r r}
\tablecolumns{4}
\tablewidth{0pt}
\tablecaption{Color-Magnitude Fits \label{table1}}
\tablehead{
\colhead{Sample} &
\colhead{No.} &
\colhead{Color$^a$} &
\colhead{Dispersion$^b$}\\
\colhead{ } &
\colhead{ } & \colhead{$U-B$} & \colhead{mag}}
\startdata
Cluster - All    & 71 & 0.438 $\pm$ 0.012 & 0.10 (0.11) \\
Cluster - Inner  & 23 & 0.462 $\pm$ 0.021 & 0.10 (0.12) \\
Cluster - Middle & 25 & 0.421 $\pm$ 0.019 & 0.09 (0.11) \\
Cluster - Outer  & 23 & 0.436 $\pm$ 0.022 & 0.10 (0.11) \\
Field            & 21 & 0.488 $\pm$ 0.029 & 0.13 (0.17) \\
\enddata
\tablecomments{\\
$^a$Colors are  from fits assuming a slope of 0, i.e., that colors are independent of luminosity,
using the method of \cite{beers90};
errors are RMS.
$^b$Dispersions are as measured; values in parentheses are the expected dispersions based
on 68\% confidence limits derived from GIM2D measurements. 
}
\end{deluxetable}

\newpage

\begin{figure}
\plotone{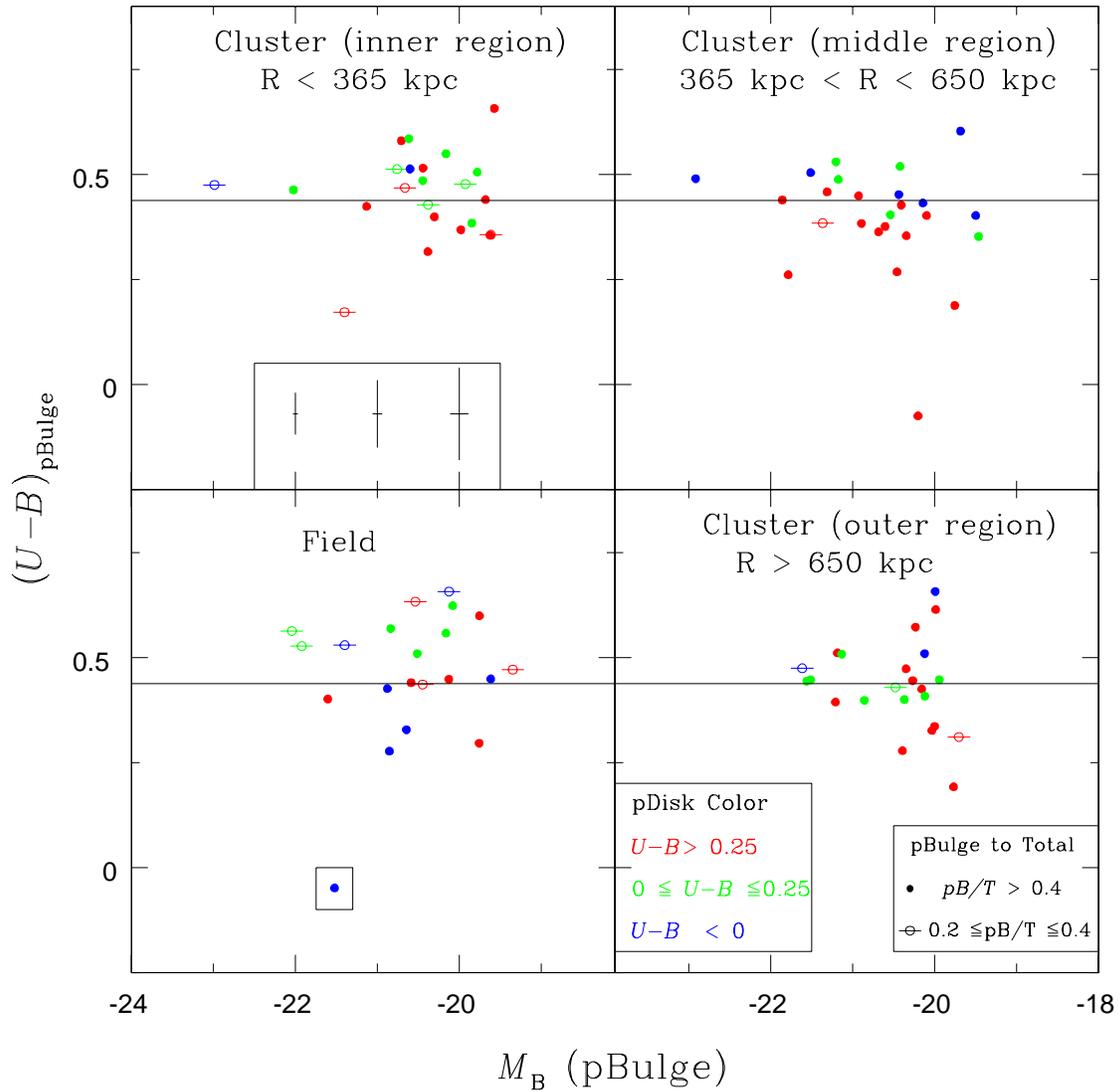}
\caption {Restframe $U-B$ color vs $M_{B}$ for GSS field and MS1054-03
cluster pbulges.  The solid line at $U-B = 0.44$ is set at
the median color of the full cluster sample. To 
show any dependence on the environment, the cluster
sample is divided into thirds by cluster-centric
distance.  
The symbols show the pB/T ratio and their colors 
indicate {\it pdisk} colors.  
The outlier in the field (boxed) has peculiar
morphology. The top-left enclosure shows
68\% confidence-limit errors in color and luminosity from GIM2D at three
luminosities.
\label{cmd}}
\vskip 0.3cm
\end{figure}


\end{document}